\documentclass[aps,twocolumn,footinbib,floatfix,showpacs]{revtex4}
\begin{document}

\title{Gaussian Dynamics is Classical Dynamics}
      \vbox to 0pt{\vss 
                    \hbox to 0pt{\hskip-40pt\rm LA-UR-02-5568\hss} 
                   \vskip 25pt} 
\preprint{LA-UR-02-5568}
\author{Salman Habib} 
\email{habib@lanl.gov}
\homepage{http://t8web.lanl.gov/people/salman/} 
\affiliation{MS B285, Theoretical Division, The University of
California, Los Alamos National Laboratory, Los Alamos, NM 87545}      
       
\date{\today} 

\begin{abstract}

A direct comparison of quantum and classical dynamical systems can be
accomplished through the use of distribution functions. This is useful
for both fundamental investigations such as the nature of the
quantum-classical transition as well as for applications such as
quantum feedback control. By affording a clear separation between
kinematical and dynamical quantum effects, the Wigner distribution is
particularly valuable in this regard. Here we discuss some
consequences of the fact that when closed-system classical and quantum
dynamics are treated in Gaussian approximation, they are in fact
identical. Thus, it follows that several results in the so-called
`semiquantum' chaos actually arise from approximating the classical,
and not the quantum dynamics. (Similarly, opposing claims of quantum
suppression of chaos are also suspect.) As a simple byproduct of the
analysis we are able to show how the Lyapunov exponent appears in the
language of phase space distributions in a way that clearly underlines
the difference between quantum and classical dynamical situations. We
also informally discuss some aspects of approximations that go beyond
the Gaussian approximation, such as the issue of when quantum
nonlinear dynamical corrections become important compared to nonlinear
classical corrections.

\end{abstract}
\maketitle

\section{Introduction and Discussion}

A formal comparison of closed classical and quantum dynamical systems
is conceptually difficult because what is natural in the classical
description (trajectories) is not what is natural in the quantum
description (amplitudes) and {\em vice versa}. While ``semiclassical''
limits of quantum mechanics are available (WKB, large $N$, coherent
states, etc.) an important obstruction is the singular nature of the
limit $\hbar\rightarrow 0$, and the fact that this limit does not
commute with the limit $t\rightarrow\infty$. Nowhere is this problem
more important than in the study of quantum chaos.

Here we adopt the viewpoint that the most natural language in which to
compare closed classical and quantum systems is the language of phase
space distribution functions as first set forth by Wigner
\cite{EW}. Since quantum mechanics is a probabilistic theory such a
description is physically attractive. Moreover, it has also turned out
to be fruitful from a calculational perspective: Among several other
applications, Wigner functions have been used to compute quantum
corrections to classical statistical mechanics \cite{IZ,OR}, to
clarify the WKB limit of quantum mechanics for both regular and
stochastic orbits \cite{MB,AV,HSC}, to study quantum dynamics
\cite{HWD}, and, in particular, dynamical aspects of quantum chaos
\cite{BB,RF1} including the effects of decoherence \cite{HSZ,ben}.
However, we do wish to emphasize that results of this paper do not
depend in any essential way on using the Wigner function: It is a
convenient device to represent the density matrix, but certainly not
necessary. We also emphasize that it is not our purpose here to
discuss the quantum-classical transition {\em per se} as would be the
case in a measured (hence, open) quantum dynamical system \cite{bhj}.

In the area of quantum chaos, the well-known work of Berry and Balazs
\cite{BB} is based on the idea of comparing classical and quantum
dynamics in phase space. These authors studied the evolution of a
semiclassical (WKB) Wigner function corresponding to a classical curve
in phase space, arguing that while at early times the Airy peak of
this distribution function tracks the time-dependent classical curve,
at later times, once the classical evolution produces structure on
sub-$\hbar$ scales in phase space, the quantum distribution is unable
to continue this tracking and there is a transition to a qualitatively
new regime. By identifying two sorts of nonlinear structures, `whorls'
(generated by stable motion) and `tendrils,' (generated by unstable
motion) Berry and Balazs conjectured that this transition would occur
on a timescale $t_c=O(\hbar^{-2/3})$ for whorls and $t_c=O(\ln
\hbar^{-1})$ for tendrils, the latter of which is the famous so-called
`log' time. It is important to note that the existence of the log time
does not necessarily imply the failure of semiclassical methods beyond
this time: later work has demonstrated that suitably improved
semiclassical methods can be successful well past the log
time~\cite{H1,HT1,OTH,PB}.

A naive comparison of Wigner functions and classical probability
distributions is not possible as Wigner functions are not positive
definite in general. This is a consequence of the fact that quantum
mechanics admits interference effects, whereas classical mechanics
does not. Nevertheless, one can set up a scheme to compare classical
and quantum dynamics in the following way: begin with an initial
distribution function that is acceptable both classically and quantum
mechanically (this condition may be relaxed to allowing all acceptable
quantum states, with or without a positive Wigner distribution) and
then, (i) solve the two different classical and quantum Liouville
equations, or equivalently, (ii) solve the corresponding infinite
hierarchies of dynamical equations for the corresponding moments. The
advantage of using the Wigner function is that, by introducing a
c-number phase space description of quantum dynamics, one can directly
compare the classical and quantum moments.

A key feature of the Wigner formalism that we exploit here is the
clear separation of kinematical and dynamical effects. Confusion has
arisen in the literature as a consequence of the failure to recognize
that $\hbar$ can occur in some places purely as a consequence of
quantum initial conditions and, in other places, as a consequence of
quantum dynamical corrections to classical dynamics. To sharpen one's
arguments, especially in areas as controversial as quantum chaos, it
is important to clearly distinguish between these two
situations. (Essentially this point, but phrased a little differently,
has also been made by Ballentine, Yang, and Zibin~\cite{BYZ}.)
The importance of this last point is further underscored by the fact
that classical and quantum mechanics have different underlying group
symmetries (symplectic vs. unitary) which accounts for the singular
nature of the limit $\hbar\rightarrow 0$~\cite{HMSSR,adsh}.

A clarification of just what is meant by such statements as ``the
classical limit'' is important both conceptually and operationally (as
emphasized in Ref.~\cite{BYZ}). The usual textbook arguments involving
wave packets and Ehrenfest's theorem attempt to compare a classical
description in the sense of a {\em single trajectory} with a quantum
analog, usually taken to be a sufficiently narrow wave packet. This is
a rather limited point of view, especially in chaotic systems, where,
due to their instability, the concept of individual trajectories is
not physically useful even classically, and where quantum wave packet
dynamics breaks down very quickly. It makes more sense, therefore, to
compare classical distribution functions, describing an {\em ensemble
of trajectories} with their appropriate quantum analogs as discussed
above. (Such a viewpoint is in fact implicit in Feynman's discussion
of the two-slit experiment \cite{RF}, a classic example of a highly
nonlinear dynamical system.) The above philosophy of quantum-classical
comparison has been implemented in some recent work on the
quantum-classical correspondence in kicked systems \cite{FE1}, the
driven pendulum \cite{DP}, and the driven anharmonic oscillator
\cite{HSZ}. Such an approach \cite{HMSSR} is also needed to analyze
experiments in quantum chaos utilizing ultracold atoms \cite{MR} in
order to unambiguously distinguish quantum from classical effects
\cite{bhjs}.

Turning now to moment expansions~\cite{mom}, the first important
point, discussed below, is that quantum dynamical effects do not arise
in moments of less than third order. In an approach utilizing moment
expansions around a centroid, this means that to first nontrivial
order in fluctuations there are {\em no} dynamical quantum effects:
This is the reason why the simple Gaussian approximation fails to
sample any aspects of quantum dynamics.

The infinite hierarchy of coupled equations for the moments is almost
impossible to deal with analytically; a variational approach is often
attempted by assuming a particular functional form for the
distribution functions, forcing thereby relations between lower and
higher order moments (alternatively, one may utilize a cumulant
expansion). A very popular approach is to assume the distribution
function to be Gaussian which implies that all odd moments vanish and
that all even moments may be expressed in terms of powers of the
second order moments.  It is useful to distinguish between three
different variants of the Gaussian approximation: The `time dependent
variational principle' (TDVP) method \cite{PAMD}, the `truncated
Gaussian approximation' (TGA) of which Heller's method \cite{HTGD} is
a non-self-consistent variant, and finally the multiple classical
trajectory version of the second technique \cite{H1}, which we will
refer to as the `multiple trajectory Gaussian approximation' (MTGA)
(although in principle the MTGA can be applied to both types of single
Gaussian approximations).  To avoid confusion, the term `Gaussian
approximation' will be used when discussing general properties of all
three variants, otherwise they will be referred to specifically.

It is easy to show that in TGA, and for the individual packets in
MTGA, no quantum dynamical corrections are included as only quadratic
order moments are considered. This is a generic problem with all
Gaussian approximations: in TDVP, while one does resum a certain
series (which in a naive moment expansion is merely truncated) this
resummation still does {\em not} incorporate terms arising from
quantum dynamics. (As a consequence, it is not obvious whether the TDVP
is any better than the truncated approximations.) For chaotic systems,
both the TDVP and TGA will lose accuracy once the initial distribution
becomes significantly non-Gaussian, which will happen at least on a
Lyapunov timescale (typically this breakdown occurs much sooner)
\cite{CDHR}. The MTGA is better behaved because a coherent sum of
Gaussians can be a much better approximation for the full distribution
function than a single Gaussian. The MTGA also keeps track of relative
quantum phases between its constituent wave packets and this is the
reason it can handle quantum revivals and other coherent effects.

What is true for Gaussian approximations in nonlinear systems is also
obviously true for linear systems where the Gaussian approximation is
exact. One consequence of this is that the well-known independent
oscillator model \cite{IO}, long considered a paradigm for quantum
Brownian motion, in fact contains only classical dynamics. All quantum
features in this model reside in the initial conditions. Once this key
fact is realized, the derivation of the Master equation can be
radically streamlined with no need to use path integral methods.
Moreover, new light is shed on the derivation of classical Langevin
equations and the interpretation of precisely what is meant by
`noise' when the dynamics is not Markovian, yet the Master equation
is local in time (A detailed discussion is given elsewhere \cite{AH}.)
 
The main point of the present paper may be stated in the following
way. Classically, or quantum mechanically, when one follows the
evolution of a `mean field' there are `fluctuation' corrections
(back-reaction) to the (classical) equations for the mean field. The
quantum moment expansion described above allows one to separate the
back-reaction into its classical and quantum components. When the
exact moment expansion is approximated by either a quadratic
truncation of the hierarchy, or by the Gaussian approximation, the
quantum back-reaction is completely neglected. The only quantum
feature still remaining is the uncertainty principle constraint on the
initial condition for the distribution function. Therefore, if one
wishes to compare the strengths of the quantum versus the classical
back-reaction, the Gaussian approximation is wholly inadequate. These
comments apply with equal force to Gaussian (Hartree) approximations
which have attained a certain popularity in studying initial value
problems in quantum field theory. Off-Gaussian corrections need to be
taken into account, as is possible in principle, e.g., in a systematic
$1/N$ expansion \cite{C1} (though this approximation has its own
problems \cite{macdh}). We will discuss some of these aspects further
below.
 
The Gaussian approximation has been applied to study chaos in quantum
systems~\cite{C2,PS}. In Ref.~\cite{C2}, the leading order $1/N$
expansion (equivalent to a quadratic TGA in that particular case) was
used to study a quantum mechanical model which is the
$(0+1)$-dimensional limit of QED. Classical chaos was shown to exist
in the dynamical equations for the correlation functions. In
Ref.~\cite{PS}, the TDVP was applied to a one-dimensional quartic
oscillator, and this system which is classically nonchaotic (in
contrast to the situation in Ref.~\cite{C2}), was shown, in the
approximation, to display chaos. In interpreting these examples of
``semiquantal'' chaos, two questions immediately arise. The first is
whether the chaos is just an artifact of the Gaussian approximation,
and the second, whether it has anything to do with quantum
dynamics. The existence of chaos in the general case of dynamical
variational approaches had been previously noted by Caurier {\em et
al}~\cite{cdop} who argued that the onset of chaos signaled the
breakdown of the approximation scheme. Motivated by the results of
Ref.~\cite{PS}, the question was also tackled by Sundaram and
Milonni~\cite{SM} using operator moments leading to a similar
conclusion. In yet another investigation, it was also demonstrated
that Gaussian approximations break down before the timescale on which
the approximate dynamics becomes chaotic \cite{CDHR}. (For a recent
study of semiquantum chaos in a different approximation, see
Ref.~\cite{ballsemi}.) The answer to the second question, given here,
is that Gaussian semiquantum chaos, even if approximate, has nothing
to do with quantum dynamics. Thus the result of Pattanayak and
Schieve~\cite{PS} is not counter-intuitive (how can ``adding'' quantum
mechanics to a nonchaotic classical system produce chaos?): it is a
consequence of the approximation, and selective resummation, of only
the classical back-reaction terms in the moment expansion. This also
rules out the heuristic explanation that their result is due to the
inclusion somehow of quantum ``noise,'' which then makes the system
chaotic.

The Gaussian approximation has also been applied by Zhang and
Feng~\cite{ZF} in an attempt to display an example of quantal
suppression of chaos. Just as we reject the reality of semiquantal
chaos, so must we reject such an interpretation of the
calculation. What has been shown is that the classical back-reaction,
in Gaussian approximation, can suppress chaos. This is not surprising
in principle as the Gaussian approximation is completely uncontrolled
and there is no guarantee that if the original system is
nonintegrable, the Gaussian dynamics also will be (the reverse is true
in semiquantum chaos where the original system can be integrable, and
the Gaussian approximation, not).

The above discussion leads naturally to another set of questions: What
are the timescales on which non-Gaussian corrections become important?
Are these timescales different for the classical and quantum
back-reaction terms? As argued in Refs.~\cite{BYZ,SM} it is certainly
reasonable to expect the TGA and TDVP to break down, for a classically
chaotic system, on a timescale set by the largest Lyapunov
exponent. Since ensemble averaged quantities do not possess chaotic
signatures like a late-time Lyapunov exponent~\cite{shkjks}, it is
also clear that any time the Gaussian approximation is chaotic, the
result is simply wrong. The timescale set by this false chaos, in the
cases where it occurs, is therefore precisely the timescale over which
one expects the classical non-Gaussian corrections to become
important.  It is not {\em a priori} obvious what will happen for the
quantum corrections, but given that the classical non-Gaussian
corrections must eliminate the chaos on a (relatively) fast timescale,
there is no dynamical imperative for these contributions to become
dominant at early times.

The concept of a crossover or `break' time from classical to quantum
dynamics in closed systems (chaotic or not) is often invoked. However,
the evidence for the utility of such a concept is surprisingly weak:
The notion of the crossover time could well depend on the quantities
one chooses to compare, and on the dynamical approximations one
makes. If, for example, the estimates of this time are based on TGA or
TDVP, then all one is computing is when the approximation breaks
down. That this may not have anything to do with how well classical
systems track quantum evolution (in a suitable coarse-grained sense),
both for chaotic and regular systems, has been pointed out in
Ref.~\cite{BYZ}. Evidence for this point of view is contained, for
example, in the work of Fox and Elston~\cite{FE1} on the kicked top
and the kicked pendulum. These authors compare the time evolution of
classical and quantum distributions, finding excellent agreement at
early times (as expected, the Gaussian approximation being accurate in
this regime), but also at late times, when a coarse-grained quantum
distribution is compared to the classical distribution. What this says
is that the low order moments can be insensitive to the nature of the
dynamics, despite the existence of nontrivial quantum dynamical
effects. Investigations utilizing high-resolution simulations have
shown that in some examples of chaotic dynamical systems, while the
classical and quantum moments do track each other very closely, there
is a small `mini-break' which occurs on a timescale longer than the
dynamical time \cite{HSZ}. Whether this `mini-break' has anything to
do with log time arguments is not yet clear, though more recent
numerical evidence appears to argue against this \cite{hs}.

The set of equations of the simple, albeit non-self-consistent, TGA
due to Heller~\cite{HTGD} provide an instructive link to the
computation of the actual classical Lyapunov exponents. Classical
mechanics allows arbitrarily strong shrinking of a Gaussian ball in
phase space around the fiducial trajectory; this shrinking is what
allows the effective Hamiltonian for Gaussian fluctuations to reduce
to that for linearized perturbations, which is just what defines the
Lyapunov exponents. In this way, even though the classical Liouville
equation is linear, the allowance of a singular trajectory limit
permits the existence of chaos. However, in the quantum case, because
of the finite value of $\hbar$, the shrinking allowed classically is
impossible beyond a certain point, and the effective Hamiltonian for
Gaussian fluctuations never becomes that for linearized perturbations.

Finally, it is important to point out that the Gaussian approximation
works extremely well in certain situations where localization of the
quantum state is guaranteed, as can happen in studying the dynamics of
observed quantum systems (quantum trajectories) (Cf. Ref.~\cite{bhj}).
In a related context, the use of a Gaussian ansatz for the Wigner
function in quantum feedback control has also been shown to be very
effective~\cite{dohjac,nanocool,atomcool}, even when the true state
can have substantial non-Gaussian features~\cite{atomcool}. While this
is not the focus of the present paper, some of the general formulae
for Gaussian states discussed here are also applicable to these,
otherwise quite distinct, situations.

The organization of this paper is as follows. In Section~II, we study
the classical and quantum Liouville equations, especially in the
context of moment expansions. In Section~III, we develop Gaussian
dynamics using moments and discuss various ramifications of the
results. We conclude in Section~IV with a short summary.

\section{Classical and Quantum Dynamics in Phase Space}
\subsection{Liouville Equations}

Throughout this paper we will deal with continuous-time Hamiltonian
dynamical systems, with Hamiltonian
\begin{equation}
H={p^2\over 2m}+V(x),                               \label{1}
\end{equation}
where the potential $V(x)$ is allowed to be explicitly time-dependent.
We take the system to be one-dimensional: extension to higher
dimensions is obvious. The classical equations of motion following
from the Hamiltonian~(\ref{1}) are entirely equivalent to the
classical Liouville equation 
\begin{equation}
{\partial f_c\over\partial t}=L_cf_c                 \label{2}
\end{equation}
for the phase space distribution function $f_c(x,p)$, and where the
classical Liouville operator,  
\begin{equation}
L_c=-{p\over m}{\partial \over\partial x}+{\partial V\over\partial
x}{\partial \over\partial p}.                       \label{3}
\end{equation}
Point trajectories can be recovered from the Liouville
equation~(\ref{2}) if one allows the distribution function to be a
delta function over phase space.

An important consequence of the classical evolution equation is the
exact conservation of the quantities (when they exist; $f$ is any
phase-space distribution function)
\begin{equation}
\sigma_n\equiv\int dxdp~f^n                      \label{3b}
\end{equation}
by the classical evolution equation~(\ref{2}), following from
Liouville's theorem. (The trivial case $n=1$ arises from the
conservation of total probability.) This will acquire more
significance when we proceed to the quantum case.

The corresponding quantum Liouville equation is written in terms of
the Wigner function \cite{EW}, which is a `half-Fourier transform'
of the density matrix $\rho(x_1,x_2)$ in the coordinate
representation. One introduces the sum and difference variables 
\begin{equation}
x=(x_1+x_2)/2,~~~~\Delta=x_1-x_2                        \label{4}
\end{equation}
and then defines the Wigner function as
\begin{equation}
f_W(x,p)={1\over 2\pi\hbar}\int\rho(x,\Delta)
\hbox{e}^{-ip\Delta/\hbar}d\Delta.                    \label{5}
\end{equation}
Details on the properties of this function may be found in
Refs.~\cite{WDF1,WDF2}. A particularly nice discussion, with special
emphasis on conceptual pitfalls and the distinction between
kinematical and dynamical aspects, has been given by
Tatarskii~\cite{VIT}. (See also, Ref.~\cite{SH1}.)

For the purposes of this paper, it is sufficient to note the
following: (i) $f_W$ is real and normalized to unity over the
$(x,p)$ phase space, (ii) $f_W$ is square integrable, with  
\begin{equation}
\sigma_2=\int dxdp~f_W^2\leq {1\over 2\pi\hbar},        \label{6}
\end{equation}
the equality holding for pure states
($2\pi\hbar\sigma_2=Tr\rho^2$). This is a {\em partial} statement of
the uncertainty principle in the Wigner formalism; localization of the
distribution to a delta function on phase space, while classically
allowed, is forbidden in quantum mechanics. (iii) The only pure state
for which $f_W$ is positive everywhere is one with the Gaussian wave
function (Hudson's theorem~\cite{HT})
\begin{equation}
\psi=\hbox{e}^{-ax^2+bx+c}                            \label{7}
\end{equation}
with $a$, $b$, $c$ complex and $Re~a > 0$. (iv) Expectation
values of operators that are either pure powers in position or pure
powers in momentum are directly given by the averages of $x^m$ or
$p^m$ taken with respect to $f_W$. Expectation values of mixed
operators like $\hat{x}^2\hat{p}^2$ cannot be obtained directly. The
Wigner formalism is associated with Weyl's rule for the ordering of
operators~\cite{WR}: the phase space average of a mixed term like
$x^np^m$ corresponds to the expectation value of a completely
symmetrized operator expression.
 
The dynamics of the Wigner function is easily obtained from the
quantum Liouville equation for the density matrix $\rho(x_1,x_2)$:
\begin{equation}
i\hbar{\partial\over\partial
t}\rho(x_1,x_2)=\left[H(x_1)-H(x_2)^*\right] 
\rho(x_1,x_2).                                         \label{8}
\end{equation}
The Wigner transform of this equation gives
\begin{equation}
{\partial f_W\over\partial
t}=L_cf_W+\sum_{\lambda~odd}{1\over\lambda!}\left({\hbar\over
2i}\right)^{\lambda-1}{\partial^{\lambda}V(x)\over\partial
x^{\lambda}}{\partial^{\lambda}f_W\over\partial p^{\lambda}} \label{9}
\end{equation}
where $\lambda\neq 1$, and with the assumption that the the potential
can be Taylor-expanded. Often this is not the case and Eqn.~(\ref{9})
becomes an integro-differential equation; we avoid this added
complication in this paper. On comparing Eqns.~(\ref{9}) and
(\ref{2}) we see that quantum and classical dynamics differ due to the
last term in the quantum Liouville equation~(\ref{9}). This apparent
separation of quantum and classical pieces in the evolution equation
is one of the original motivations for the Wigner function
approach. However, since the definition of the Wigner function itself
involves $\hbar$, the actual situation is not as straightforward as it
appears to be. In particular, the `quantum correction' in
Eqn.~(\ref{9}) is not unitary. (Neither is $L_c$, but their sum
is. Ref.~\cite{HMSSR} contains a further discussion including the
violation of positivity in the classical approximation.)

The quantum evolution equation~(\ref{9}) exactly conserves the
uncertainty principle constraint~(\ref{6}). This has a one-to-one
correspondence with the classical evolution preserving the
corresponding classical quantity~$\sigma_2$ of Eqn.~(\ref{3b}). It is
important to note that since the quantum evolution does not obey
Liouville's theorem, in this case, $\dot{\sigma}_n\neq 0, \forall
n\neq 1,2$. Thus, since the conservation of $\sigma_1$ is trivial in
both classical and quantum evolutions, the constancy of only
$\sigma_2$ in the quantum case acquires added significance.

\subsection{The Moment Hierarchies}

In the moment hierarchy approach to dynamics one attempts to replace a
knowledge of the distribution function by a knowledge of all-order
values of the quantities $\left\langle{x^n}\right\rangle$,
$\left\langle{p^n}\right\rangle$, and
$\left\langle{x^np^m}\right\rangle$ where the bracket denotes an
average taken over the distribution function. From now on
$\left\langle{~}\right\rangle_c$ will denote an expectation value
taken with respect to a classical distribution, and
$\left\langle{~}\right\rangle_q$ will denote an expectation value
taken with respect to the quantum distribution function. Below we
present an elementary derivation of the classical and quantum moment
equations using the evolution equations for the corresponding
distributions (for alternative approaches, see Ref.~\cite{mom}).

With the assumption that the distribution function and its derivatives
are zero at the phase space boundary, in the classical case [using the
classical Liouville equation~(\ref{2})] it is easy to show that :
\begin{eqnarray}
{d\over dt}\left\langle{x^n}\right\rangle_c&=&\int dxdp~x^n{\partial
f_c\over\partial t}   \nonumber\\
&=&{n\over m}\left\langle{x^{n-1}p}\right\rangle_c,\label{10}\\ 
{d\over dt}\left\langle{p^n}\right\rangle_c&=&\int dxdp~p^n{\partial
f_c\over\partial t}\nonumber\\
&=&-n\left\langle{p^{n-1}{\partial V\over\partial
x}}\right\rangle_c, \label{11}\\ 
{d\over dt}\left\langle{x^np^k}\right\rangle_c&=&\int
dxdp~x^np^k{\partial f_c\over\partial t}     \nonumber\\
&=&{n\over m}\left\langle{x^{n-1}p^{k+1}}\right\rangle_c\nonumber\\
&& -k\left\langle{x^np^{k-1}{\partial V\over\partial
x}}\right\rangle_c. \label{12} 
\end{eqnarray}
Of course, these equations are nothing but the Hamilton equations
of motion averaged over the initial conditions.
  
The quantum hierarchy is equally straightforward to obtain. The
evolution of the spatial moments is given by
\begin{eqnarray}
{d\over dt}\left\langle{x^n}\right\rangle_q&=&\int dxdp~x^n{\partial
f_W\over\partial t} \nonumber\\
&=&{n\over m}\left\langle{x^{n-1}p}\right\rangle_q,
\label{13}
\end{eqnarray}
formally the same as the classical case. However,
\begin{eqnarray}
{d\over dt}\left\langle{p^n}\right\rangle_q&=&\int dxdp~p^n{\partial
f_W\over\partial t}                                \nonumber\\
&=&-n\left\langle{p^{n-1}{\partial V\over\partial
x}}\right\rangle_q\nonumber\\ 
&&+\sum_{\lambda~odd}^{n,~n-1}\Theta(n,\lambda)\left\langle
{p^{n-\lambda}{\partial^{\lambda}V\over\partial
x^{\lambda}}}\right\rangle_q,                      \label{14}  
\end{eqnarray}
where
\begin{equation}
\Theta(n,\lambda)\equiv{(-1)^{\lambda}\over\lambda!}{n!\over
(n-\lambda)!}\left({\hbar\over 2i}\right)^{\lambda-1}, \label{14b} 
\end{equation}
differs from the corresponding classical equation. The upper limit of
the sum in (\ref{14}) is $n$, if $n$ is odd, and $n-1$ if $n$ is
even. Clearly the first dynamical quantum correction comes in only at
third order, where
\begin{equation}
{d\over
dt}\left\langle{p^3}\right\rangle_q=-3\left\langle{p^2{\partial
V\over\partial x}}\right\rangle_q+{\hbar^2\over 
4}\left\langle{{\partial^3V\over\partial x^3}}\right\rangle_q.
                                                   \label{15} 
\end{equation}
Finally, the cross moments evolve according to
\begin{eqnarray}
{d\over dt}\left\langle{x^np^k}\right\rangle_q&=&\int
dxdp~x^np^k{\partial f_W\over\partial t}           \nonumber\\
&=&{n\over m}\left\langle{x^{n-1}p^{k+1}}\right\rangle_q-k\left
\langle{x^np^{k-1}{\partial V\over \partial 
x}}\right\rangle_q                                  \nonumber\\
&& +\sum_{\lambda~odd}^{k,k-1}\Theta(k,\lambda)\left
\langle{x^np^{k-\lambda}{\partial^{\lambda} V\over\partial
x^{\lambda}}}\right\rangle_q.                        \label{16} 
\end{eqnarray} 
As in Eqn.~(\ref{14}), the upper limit of the sum is $k$ if $k$ is
odd, and $k-1$, if it is even.

A moment expansion around the centroid can now be developed by
expanding in `fluctuations' around `mean fields,' 
\begin{eqnarray}
x&=&\left\langle{x}\right\rangle_c+\delta,            \label{17}\\
p&=&\left\langle{p}\right\rangle_c+\eta,              \label{18}
\end{eqnarray} 
in the classical case, and
\begin{eqnarray}
x&=&\left\langle{x}\right\rangle_q+\delta,             \label{19}\\
p&=&\left\langle{p}\right\rangle_q+\eta,                \label{20}
\end{eqnarray}
in the quantum case. It is clear from the definition of the
fluctuations that their expectation values vanish identically. 

The strategy to follow now is simple: One begins with the lowest order
classical dynamical equations from (\ref{10})-(\ref{12}), or the
corresponding quantum equations (\ref{14})-(\ref{16}), and substitutes
in the appropriate pair from equations (\ref{17})-(\ref{20}) given
above. One then finds terms involving expectation values of higher
order fluctuations. Dynamical equations for these quantities can then
be found by going to higher order in (\ref{10})-(\ref{12}) or
(\ref{17})-(\ref{20}). Care has to be taken in interpreting the moment
expansion: the classical and quantum mean field equations are not
identical and the fluctuations in the quantum case contain both
`classical' and `quantum' contributions. An analogous moment expansion
was performed by Sundaram and Milonni~\cite{SM} by working in a
Heisenberg operator formalism. The moment hierarchy described below is
completely equivalent to the set of equations they derived. The
advantage here is that we can compare directly with the corresponding
classical hierarchy.

To illustrate the above procedure we now carry out the moment
expansion to quadratic order in the fluctuations. As will be seen
below, to this order, there is no difference between the classical and
quantum hierarchies (this is well-known in quantum optics, see, e.g.,
Ref.~\cite{FR}) so it is sufficient merely to present the classical
situation. The equations for the average quantities are [set $n=1$ in
(\ref{10}) and (\ref{11})]: 
\begin{eqnarray}
{d\over
dt}\left\langle{x}\right\rangle_c&=&{\left\langle{p}\right\rangle_c\over
m},                                              \label{21}\\ 
{d\over dt}\left\langle{p}\right\rangle_c&=&-\left\langle{{\partial
V\over\partial x}}\right\rangle_c     \nonumber\\ 
&=&-V^{(1)}_c-{1\over 2!} \left\langle{\delta^2}\right\rangle_c
V^{(3)}_c+\cdots,                                 \label{22} 
\end{eqnarray}
where
\begin{equation}
V^{(n)}_c\equiv\left[{\partial^n V(x)\over \partial
x^n}\right]_{x=\left\langle{x}\right\rangle_c}.         \label{23}
\end{equation}
To obtain a closed set of equations we need to know the equations for
the fluctuations correct to quadratic order. We begin by noting that
\begin{equation}
\left\langle{x^2}\right\rangle_c=\left\langle{x}\right
\rangle_c^2+\left\langle{\delta^2}\right\rangle_c      \label{24} 
\end{equation}
which implies directly in turn
\begin{equation}
{d\over dt}\left\langle{x^2}\right\rangle_c=2\left\langle{x}\right
\rangle_c{d\over dt}\left\langle{x}\right\rangle_c+{d \over
dt}\left\langle{\delta^2}\right\rangle_c.                \label{25}
\end{equation}
But we know already from the hierarchy that [set $n=2$ in (\ref{10})],
\begin{eqnarray}
{d\over dt}\left\langle{x^2}\right\rangle_c&=&{2\over
m}\left\langle{xp}\right\rangle_c         \nonumber\\ 
&=&{2\over m}\left(\left\langle{x}\right\rangle_c\left\langle{p}
\right\rangle_c+\left\langle{\delta\eta}\right\rangle_c\right)
\nonumber\\ 
&=&2\left\langle{x}\right\rangle_c{d\over
dt}\left\langle{x}\right\rangle_c+{2\over
m}\left\langle{\delta\eta}\right\rangle_c.           \label{26}
\end{eqnarray}
Comparing (\ref{25}) and (\ref{26}), we find
\begin{equation}
{d\over dt}\left\langle{\delta^2}\right\rangle_c={2\over
m}\left\langle{\delta\eta}\right\rangle_c      \label{27} 
\end{equation}
which is the first of the required equations. Notice that the
derivation of this equation is in fact exact as no terms were
neglected. This is not true for the other two equations to which we
now pay attention. Proceeding as above, we first write
\begin{equation}
{d\over dt}\left\langle{p^2}\right\rangle_c=2\left\langle{p}\right
\rangle_c{d\over dt}\left\langle{p}\right\rangle_c+{d \over 
dt}\left\langle{\eta^2}\right\rangle_c.                 \label{28} 
\end{equation}
then setting $n=2$ in (\ref{11}), we find
\begin{eqnarray}
{d\over
dt}\left\langle{p^2}\right\rangle_c&=&-2\left\langle{p{\partial
V\over\partial x}}\right\rangle_c \nonumber\\ 
&=&-2\left[\left\langle{p}\right\rangle_cV^{(1)}_c+{1\over
2!}\left\langle{p}\right\rangle_c
\left\langle{\delta^2}\right\rangle_c V^{(3)}_c\right.\nonumber\\ 
&&\left.+\left\langle{\eta\delta}\right\rangle_cV^{(2)}_c+\cdots
\right]                           \nonumber\\    
&=&2\left\langle{p}\right\rangle_c{d\over
dt}\left\langle{p}\right\rangle_c-2\left\langle{\eta\delta}
\right\rangle_cV^{(2)}_c +\cdots.            \label{29} 
\end{eqnarray}
Comparing (\ref{28}) and (\ref{29}), we obtain
\begin{equation}
{d\over dt}\left\langle{\eta^2}\right\rangle_c=-2\left\langle
{\eta\delta}\right\rangle_cV^{(2)}_c.       \label{30}
\end{equation}
It is now apparent how to proceed to get the last equation and we
merely write the final result
\begin{equation}
{d\over dt}\left\langle{\eta\delta}\right\rangle_c={\left\langle
{\eta^2}\right\rangle_c\over m}-\left\langle{\delta^2}
\right\rangle_cV^{(2)}_c.                   \label{31} 
\end{equation}
Equations (\ref{21}), (\ref{22}), (\ref{27}), (\ref{30}), and
(\ref{31}) form a closed set of dynamical equations. Since at most
only second order moments were used in the derivation, these five
classical equations are formally identical to their quantum
counterparts. Had we truncated the moment expansion at one higher
order there would have been a difference between the quantum and
classical equations since, as we have seen, the first quantum
dynamical correction comes in only at third order. 

General expressions can be written for the exact classical and quantum
moment hierarchies by extending the procedure outlined above. In the
classical case, restricting attention only to those moments relevant
to future discussion of the Gaussian approximation, one finds,
\begin{eqnarray}
{d\over dt}\left\langle{\delta^n}\right\rangle_c&=&{n\over
m}\left\langle{\delta^{n-1}\eta}\right\rangle_c,   \label{32}\\
{d\over dt}\left\langle{\eta^n}\right\rangle_c&=&n\left[
\left\langle{\eta^{n-1}}\right\rangle_c\left\langle{{\partial 
V\over\partial x}}\right\rangle_c \right.\nonumber\\
&&\left.- \left\langle{\eta^{n-1}{\partial V\over\partial
x}}\right\rangle_c\right],                        \label{33}\\ 
{d\over dt}\left\langle{\delta^n\eta}\right\rangle_c&=&\left\langle
{\delta^n}\right\rangle_c\left\langle{{\partial V\over\partial
x}}\right\rangle_c                                 \nonumber\\  
&&- \left\langle{\delta^n {\partial V\over\partial x}}\right\rangle_c
+ {n\over m}\left\langle{\delta^{n-1}\eta^2}\right\rangle_c. \label{34} 
\end{eqnarray}        

The quantum hierarchy is only slightly more complicated. For the same
moments considered above, 
\begin{eqnarray}
{d\over dt}\left\langle{\delta^n}\right\rangle_q&=&{n\over
m}\left\langle{\delta^{n-1}\eta}\right\rangle_q,  
                                                       \label{35}\\
{d\over dt}\left\langle{\eta^n}\right\rangle_q&=&n\left[\left\langle
{\eta^{n-1}}\right\rangle_q\left\langle{{\partial V\over\partial
x}}\right\rangle_q \right.                      \nonumber\\ 
&&\left.- \left\langle{\eta^{n-1}{\partial V\over\partial
x}}\right\rangle_q\right] + \Delta^{(n)},           \label{36}\\ 
{d\over dt}\left\langle{\delta^n\eta}\right\rangle_q&=&\left\langle
{\delta^n}\right\rangle_q\left\langle{{\partial V\over\partial
x}}\right\rangle_q - \left\langle{\delta^n {\partial V\over\partial
x}}\right\rangle_q                               \nonumber\\ 
&& + {n\over m}\left\langle{\delta^{n-1}\eta^2}\right\rangle_q,
                                               \label{37} 
\end{eqnarray}
where
\begin{equation}
\Delta^{(n)}=\sum_{\lambda~odd}^{n,n-1}\Theta(n,\lambda)
\left\langle{\left(\left\langle{p}\right\rangle_q+
\eta\right)^{n-\lambda} {\partial^{\lambda}V\over\partial
x^{\lambda}}}\right\rangle_q.                     \label{38}
\end{equation}
Note that among the equations displayed, only (\ref{36}) contains
quantum corrections compared to the classical equations. This is
because (\ref{13}) is already formally identical to the corresponding
classical equation, as is (\ref{16}) in the case $k=1$.

\section{The Gaussian Approximation}
\subsection{Time Dependent Variational Principle}

We now proceed to the first variant of the Gaussian approximation. The
idea here is to (i) {\em assume} that the distribution function is at
all times a Gaussian (for more mathematical details on the formal
properties of such states, see Refs.~\cite{RJL,BS}) and then, (ii) to
write down the corresponding dynamical equations for the moment
hierarchy. The approach given here is a generalization of the time
dependent variational principle first considered by Dirac~\cite{PAMD}
by considering mixed as well as pure states. The moment hierarchy
approach is far more convenient than a direct application of a
variational principle as carried out, for example, in
Ref.~\cite{EJP}. As will be apparent in what follows, this
approximation only samples the moment hierarchy to quadratic order,
and hence contains no dynamical quantum corrections.

We assume the system to be always described by a general Gaussian
distribution (here we make no distinction between classical and
quantum) of the form
\begin{eqnarray} 
f={1\over N}\exp
&[&-a(x-\bar{x}(t))^2-b(p-\bar{p}(t))^2 \nonumber\\
&&+c(x-\bar{x}(t))(p-\bar{p}(t))], \label{39} 
\end{eqnarray} 
or in a more compact notation:
\begin{equation}
f={1\over\sqrt{Det(2\pi S)}}\exp\left[-{1\over
2}\sum_{j,k}(x_j-\bar{x}_j)S^{-1}_{jk}(x_k-\bar{x}_k)\right],
\label{39b}
\end{equation}
with $x_1=x$, $x_2=p$, and the matrices,
\begin{equation}
S^{-1}=\left(\begin{array}{cc}
2a & -c\\
-c  & 2b
\end{array}\right),      \label{39c}
\end{equation}
\begin{equation}
S={1\over 4ab-c^2}\left(\begin{array}{cc}
2b & c\\
c  & 2a
\end{array}\right).      \label{39d}
\end{equation}
The distribution is normalized to unity over phase space, i.e., 
\begin{equation}
N=\sqrt{Det(2\pi S)}={2\pi\over\sqrt{4ab-c^2}}, \label{40} 
\end{equation}
and satisfies the constraint (\ref{6}). This extra condition in turn
leads to a constraint on the parameters of the distribution:
\begin{equation} 
4ab-c^2=(4\pi\sigma_2)^2.       \label{41} 
\end{equation} 
This purely kinematical condition must be satisfied for quantum
distributions. In general, classical distributions need not satisfy
this condition, but if they do, then, as already discussed, this
condition is exactly preserved by the classical dynamics.

Given the distribution function specified above, it is trivial to show
that
\begin{equation}
\left\langle{x}\right\rangle=\bar{x},~~~~\left\langle{p}\right
\rangle=\bar{p},                      \label{42} 
\end{equation}
and that the quadratic moments are
\begin{eqnarray}
\left\langle{\delta^2}\right\rangle=S_{11}&=&{1\over
2\alpha},~~~~\alpha=a-{c^2\over 4b}={4\pi^2\sigma_2^2\over b},
                                        \label{46}\\ 
\left\langle{\eta^2}\right\rangle=S_{22}&=&{1\over
2\beta},~~~~\beta=b-{c^2\over 4a}={4\pi^2\sigma_2^2\over a},
                                         \label{47}\\ 
\left\langle{\eta\delta}\right\rangle=S_{12}&=&{c\over
4a\beta}={c\over (4\pi\sigma_2)^2}.   \label{48} 
\end{eqnarray}
The kinematical constraint (\ref{41}) can now be expressed as
\begin{equation}
\left\langle{\eta^2}\right\rangle\left\langle{\delta^2}
\right\rangle=\left\langle{\eta\delta}\right\rangle^2+
{1\over(4\pi\sigma_2)^2},                \label{49}
\end{equation}
which is a statement of the uncertainty principle generalized to mixed
states. For pure states, $\sigma_2=1/2\pi\hbar$, and (\ref{49}) takes the
familiar form (the wave function is now that of some Gaussian state,
more general than just a squeezed coherent state)
\begin{equation}
\left\langle{\eta^2}\right\rangle\left\langle{\delta^2}\right
\rangle=\left\langle{\eta\delta}\right\rangle^2+{\hbar^2\over 
4}.                                         \label{49a} 
\end{equation}
All odd-order moments are zero and the higher order even moments can
be computed from the quadratic ones by a simple application of the
result, valid for Gaussian distributions,
\begin{equation}
\left\langle{(x_j-\bar{x}_j)g({\bf x})}\right\rangle=\sum_k
S_{jk}\left\langle{{\partial g({\bf 
x})\over \partial x_k}}\right\rangle,                \label{43a}
\end{equation}
where ${\bf x}=(x_1,x_2)$ and $g({\bf x})$ is some arbitrary function
of ${\bf x}$. The moments relevant to us are
\begin{eqnarray}
\left\langle{\delta^{2n}}\right\rangle&=&{(2n-1)!\over
2^{n-1}(n-1)!}\left\langle{\delta^2}\right\rangle^n,  \label{43}\\
\left\langle{\eta^{2n}}\right\rangle&=&{(2n-1)!\over
2^{n-1}(n-1)!}\left\langle{\eta^2}\right\rangle^n,    \label{44}\\
\left\langle{\eta\delta^{2n+1}}\right\rangle&=&{(2n+1)!\over 2^n
n!}\left\langle{\eta\delta}\right\rangle\left\langle
{\delta^2}\right\rangle^n,                            \label{45}
\end{eqnarray}
where (\ref{49}) has been used to simplify the last expression. The
above expressions are the only ones needed to obtain the moment
hierarchy in Gaussian approximation.   

The dynamical equations under the Gaussian approximation are easy to
obtain. First,
\begin{equation}
{d\over dt}\left\langle{x}\right\rangle={\left\langle{p}\right\rangle
\over m}                                       \label{49b} 
\end{equation}
or
\begin{equation}
{d\over dt}\bar{x}={\bar{p}\over m}              \label{49c}
\end{equation}
and
\begin{eqnarray}
{d\over dt}\left\langle{p}\right\rangle={d\over
dt}\bar{p}&=&-\left\langle{\partial V\over\partial x}\right\rangle          
                                                       \nonumber\\
&=&-V^{(1)}-{\left\langle{\delta^2}\right\rangle\over
2!}V^{(3)}-{\left\langle{\delta^3}\right\rangle\over 
3!}V^{(4)}-\cdots                                      \nonumber\\
&=&-\sum_{n=0}^{\infty}{\left\langle{\delta^2}\right\rangle^n\over
n!2^n}V^{(2n+1)}                                  \label{50}
\end{eqnarray}
where we have used (\ref{43}) in the last step. The dynamical equation
for $\left\langle{\delta^2}\right\rangle$ is nothing but (\ref{27})
which is already exact. The equation for
$\left\langle{\eta^2}\right\rangle$ follows from (\ref{33}) or
(\ref{36}):  
\begin{eqnarray}
{d\over dt}\left\langle{\eta^2}\right\rangle&=&-2\left\langle
{\eta{\partial V\over\partial x}}\right\rangle\nonumber\\ 
&=&-2\left[\left\langle\eta\left(V^{(1)}+\delta V^{(2)}+{\delta^2
\over 2!}V^{(3)}\right.\right.\right.                  \nonumber\\
&& \left.\left.\left.+{\delta^3\over
3!}V^{(4)}+\cdots\right)\right\rangle\right] 
                                                       \nonumber\\ 
&=&-\sum_{k=0}^{\infty}{V^{(2k+2)}\over
2^{k-1}k!}\left\langle{\eta\delta}\right\rangle\left\langle
{\delta^2}\right\rangle^{k}.                   \label{51}   
\end{eqnarray}
This result has been written in a simple form by using (\ref{45})
along with the constraint (\ref{49}). The last equation we have to
deal with is for $\left\langle{\delta\eta}\right\rangle$, this
following from (\ref{34}): 
\begin{eqnarray}
{d\over dt}\left\langle{\delta\eta}\right\rangle&=&{1\over
m}\left\langle{\eta^2}\right\rangle-\left\langle{\delta{\partial
V\over \partial x}}\right\rangle                 \nonumber\\
&=&{1\over m}\left\langle{\eta^2}\right\rangle-\left\langle
\delta\left(V^{(1)}+\delta 
V^{(2)}+{\delta^2\over 2!}V^{(3)}\right.\right.   \nonumber\\
&&\left.\left. +{\delta^3\over 3!}V^{(4)}+\cdots\right)\right\rangle   
                                                 \nonumber\\
&=&{1\over m}\left\langle{\eta^2}\right\rangle-\sum_{n=0}^{\infty}
{\left\langle{\delta^2}\right\rangle^{n+1}\over 2^n n!}V^{(2n+2)}.
                                               \label{52} 
\end{eqnarray}
Note that we have used (\ref{43}) in the last step. 

We now have five equations, two for the mean fields, and three for the
fluctuations. It is easy to check that the constraint (\ref{49a}) is
exactly preserved by the dynamical equations. Therefore the constraint
may be used to eliminate one of the three dynamical equations for the
fluctuations. For example, if we choose to keep
$\left\langle{\delta^2}\right\rangle$ and
$\left\langle{\delta\eta}\right\rangle$ as the preferred variables,
then the two dynamical equations for the fluctuations are (\ref{27})
and (\ref{52}) with $\left\langle{\eta^2}\right\rangle$ written in
terms of $\left\langle{\delta^2}\right\rangle$ and
$\left\langle{\delta\eta}\right\rangle$ using (\ref{49}). Notice that
these equations were obtained without having to go to moments higher
than those of order two. Therefore they do not contain any quantum
dynamical contributions, just as advertized. Of course, $\hbar$ can
still appear in these equations, but in a purely kinematical
sense. This will happen if the initial state is pure
($\sigma_2=1/2\pi\hbar$) and we use the constraint to eliminate one of
the dynamical equations for the fluctuations.

To conclude our discussion of the TDVP, we note that this approximation
is conservative and further, can be mapped to a classical Hamiltonian
for a new set of dynamical variables~\cite{RM}. The first point can be
easily checked by writing the expectation value of the Hamiltonian
(here assumed explicitly time independent) in Gaussian approximation, 
\begin{eqnarray}
\left\langle{H}\right\rangle&=&{1\over
2m}\left\langle{p^2}\right\rangle+\left\langle{V(x)}\right\rangle
\nonumber\\ 
&=&{1\over 2m}\left[\left\langle{p}\right\rangle^2+\left\langle
{\eta^2}\right\rangle\right]+\sum_{k=0}^{\infty} 
{V^{(2k)}\over 2^k k!}\left\langle{\delta^2}\right\rangle^k \label{53} 
\end{eqnarray}
and then verifying, using the dynamical equations given above, that
\begin{equation}
{d\over dt}\left\langle{H}\right\rangle=0.              \label{54}
\end{equation}
It is interesting to note that any consistent truncation of the
TDVP is also conservative (by consistent truncation, we mean: keep
only the moments up to some finite order, i.e., go up to the
{\em same} finite order in all the equations). As it happens, Heller's
approximation~\cite{HTGD} is not consistent in this sense because it
does not account for the back-reaction on the mean field thereby
missing a quadratic term that should have been kept; consequently it
fails to be conservative. 

It is easy to show that the four dynamical equations for the centroid
and two chosen fluctuation variables are in fact Hamiltonian.
Eliminating $\left\langle{\eta^2}\right\rangle$ from (\ref{52}) using
the constraint (\ref{49}) and introducing the new
variables~\cite{RM}\cite{PS}
\begin{equation}
\left\langle{\delta^2}\right\rangle\equiv\rho^2~;~~~~\left\langle
{\eta\delta}\right\rangle\equiv\rho\gamma,     \label{541} 
\end{equation}
in (\ref{53}), we find that the Hamiltonian
\begin{eqnarray}
H_G=\left\langle{H}\right\rangle&=&{1\over 2m}\bar{p}^2 + {1\over
2m}\gamma_0^2 + {1\over 2m\rho^2(4\pi\sigma_2)^2}   \nonumber\\
&& + \sum_{n=1}^{\infty}{\rho^{2n}\over 2^n
n!}V^{(2n)}(\bar{x})                                \label{542}
\end{eqnarray}
generates the correct equations of motion as can be easily
verified. It follows trivially from (\ref{542}) that any consistent
TGA is also Hamiltonian.  

The expectation value of the Hamiltonian (\ref{53}) is precisely the
quantity that is minimized in the Gaussian effective potential
approach~\cite{GEP}. Since $\hbar$ appears only in kinematical guise
in (\ref{53}) it follows that no essential quantum effects (i.e.,
those without formal classical analogs) can be seen in this
approximation. For example, tunneling/overlap corrections in a
double-well system are not accessible in simple Gaussian
approximation~\cite{AHK}.
 
\subsection{Semiquantum Chaos and Related Issues}

In Ref.~\cite{PS}, Pattanayak and Schieve studied a one-dimensional
anharmonic oscillator in Gaussian approximation. (See also
Ref.~/cite{C2}.) Even though the original system is clearly not
chaotic, in the approximation it can be, because we now have four
coupled nonlinear dynamical equations, rather than the original
two. This is the main technical point of their paper and is, of
course, correct. The authors stressed the surprising nature of their
result, that dynamics of expectation values may be chaotic, ``even
though the system has regular classical behavior and the quantum
behavior has been assumed regular''~\cite{PS}. Their conclusion has
been criticized by Sundaram and Milonni~\cite{SM} who point out that
the Gaussian approximation becomes unreliable very quickly when either
the classical system is chaotic, or when the Gaussian dynamical
equations are themselves chaotic. The more general theoretical setting
is the following: Any variational approximation to the dynamics of a
quantum system based on the Dirac action principle leads to a
classical Hamiltonian dynamics for the variational
parameters~\cite{kk}. Since this Hamiltonian is generically nonlinear
and nonintegrable, the dynamics thus generated can be chaotic, in
distinction to the exact quantum evolution~\cite{cdop,CDHR}; the
Gaussian approximation represents a particularly simple special case.

Once we realize that, in fact, the analysis presented in
Refs.~\cite{C2} and \cite{PS} is purely classical, the results are
easy to understand.  First, they are not surprising in the sense that
quantum mechanics is {\em not} somehow catalyzing chaos. Second, we
see why the approximation breaks down as it does: if the original
system is chaotic then, `tendrilization,' in the sense of Berry and
Balazs~\cite{BB} will happen on a timescale set by the maximal
Lyapunov exponent by which time the Gaussian approximation is
definitely wrong.  If, on the other hand, the original classical
system is not chaotic, but the Gaussian set of equations are, all this
tells us is that the non-Gaussian corrections are becoming very
important on the timescale set by the maximal Lyapunov exponent of the
false chaos, and need to be included to obtain the correct classical
dynamics. Third, the explanation that quantum noise is somehow
promoting the chaos is not tenable since the Gaussian approximation
contains only classical fluctuations.

An ostensibly opposite result was obtained by Zhang and
Feng~\cite{ZF}: These authors implemented the TDVP approach for the
kicked rotor, finding a suppression of the classical chaos. This, they
claimed, was a simple example of a generic suppression of classical
chaos by quantum mechanics. Our interpretation of their result is
different: since it is a consequence of merely classical dynamics,
what they have really displayed is a suppression of chaos by the
classical back-reaction, computed in Gaussian approximation. Thus,
there really is no information about quantum suppression (or even of a
classical `fluctuation suppression,' since the calculation is only
approximate). Real quantum suppression of classical chaos in this
model is discussed in Refs.~\cite{CIS,FGP}.

\subsection{The Truncated Gaussian Approximation}

The full Gaussian approximation (TDVP) involves summations to all
orders of the fluctuations (though these are somewhat trivial as the
higher terms that are summed are disconnected pieces, i.e., no
cumulants of order higher than two are encountered). This
approximation differs from a truncation approach where one merely
truncates the moment hierarchy at some finite order. We can
incorporate the truncation idea in Gaussian approximation by requiring
the distribution to be Gaussian but still stopping the summations at
some finite order. As discussed above, this truncation is conservative
and Hamiltonian in the sense that the corresponding equations of
motion for dynamical variables can be derived from a Hamiltonian. The
TGA may be useful in some situations because of its computational
simplicity.

An example of this idea is the approach due to Heller~\cite{HTGD}. In
our language, Heller's instructions are the following: (i) Drop the
fluctuation corrections to the equation for the average momentum
(\ref{50}), and (ii) keep only the leading order (quadratic)
corrections in the equations for the moments [Eqns. (\ref{27}),
(\ref{51}), and (\ref{52})]. The motivation for this approach as
originally stated is to provide a semiclassical limit that does not
quite correspond to the usual $\hbar\rightarrow 0$ limit of WKB theory
(For example, the quadratic order TGA is exact for the harmonic
oscillator, but WKB is not). As will become clear in what follows,
dynamically this is still a classical approach and its ostensible
quantum features are easily explained: (i) the appearance of $\hbar$
is purely kinematical, and (ii) ``penetration'' of the Gaussian into
classically forbidden territory is simply a classical consequence of
the distribution function having support at high momenta. It is no
surprise that this approximation has trouble with tunneling (which is
accessible in a WKB approach) since it does not incorporate quantum
dynamical corrections.

By carrying through with the prescription given above, we obtain the
dynamical equations:  
\begin{eqnarray}
{d\over dt}\bar{x}&=&{\bar{p}\over m},                \label{53b}\\
{d\over dt}\bar{p}&=&-V^{(1)},                        \label{54b}\\
{d\over dt}\left\langle{\delta^2}\right\rangle&=&{2\over
m}\left\langle{\delta\eta}\right\rangle,             \label{55}\\ 
{d\over dt}\left\langle{\delta\eta}\right\rangle&=&{1\over
m}\left\langle{\eta^2}\right\rangle-\left\langle{\delta^2}
\right\rangle V^{(2)},                              \label{56}\\
{d\over dt}\left\langle{\eta^2}\right\rangle&=&-2\left\langle
{\delta\eta}\right\rangle V^{(2)}.                 \label{57}  
\end{eqnarray}
The first two equations (\ref{53}) and (\ref{54}) are just the
classical equations of motion for a particle in the potential $V(x)$.
This approximation thus ignores the back-reaction of fluctuations on
the mean field. As for the fluctuations themselves, they are sensitive
only to the second derivative of the potential, as in the quadratic
approximation of Section IIB. Heller's method is not equivalent to a
quadratic truncation because there is no back-reaction on the mean
fields and because it still enforces the constraint (\ref{49}),
reducing the number of dynamical equations from five to four. However,
it is not a consistent quadratic TGA because of the failure to include
the back-reaction. Just as in the TDVP and quadratic approximations,
this approximation only includes classical dynamics.

As a consequence of the lack of self-consistency at quadratic order,
the expectation value of the energy is not conserved: 
\begin{equation}
{d\over dt}{\left\langle{H}\right\rangle}={\left\langle{p}\right
\rangle\over 2m}V^{(3)}\left\langle{\delta^2}
\right\rangle.                                      \label{58}    
\end{equation}
It is a trivial matter to improve this approximation by keeping the
back-reaction consistently to quadratic order. (Note, however, that
this approximation is conservative for linear systems.) Even though
this method is not unitary, it has been applied very successfully,
especially in its multiple trajectory variant, to chaotic
systems~\cite{H1,HT1}.

The implementation of the TGA in the Schr\"odinger picture also
involves taking into account a time varying global (Maslov) phase for
the wave function. Since the Wigner function is a transform of the
density matrix, it does not contain this phase information. If there
was more than one packet (as in MTGA), however, the Wigner function
would keep relative phase information for which there is no classical
analog.

\subsection{Connection with Lyapunov Exponents}

A straightforward observation is that the set of equations (\ref{53b})
- (\ref{57}) are exactly those needed to determine the classical
Lyapunov exponent, with $(\bar{x},\bar{p})$ specifying a fiducial
classical trajectory, and the fluctuation variables specifying the
linearized evolution of a Gaussian ball of initial conditions around
the fiducial trajectory. As before, imposition of the constraint
arising from the constancy of $\sigma_2$ leaves only two equations for
the fluctuations, and these are Hamiltonian separately, with
\begin{equation}
H_F={1\over 2m}\gamma^2 + {1\over 2m\rho^2(4\pi\sigma_2)^2} + {1\over
2}V^{(2)}(\bar{x})\rho^2.                         \label{59}
\end{equation}
This would be precisely the linearized Hamiltonian for the system were
it not for the second term which enforces the constraint. The
definition of the Lyapunov exponent requires the double limit of first
squeezing the initial Gaussian ball onto the fiducial trajectory (in
{\em all} phase space directions) and then taking the limit
$t\rightarrow\infty$. The first limit corresponds to taking
$\sigma_2\rightarrow\infty$ (clearly not allowed quantum mechanically
for $\hbar$ finite), which is the same as dropping the second term in
$H_F$. Incidentally, this is also a good way to see how chaos can
occur in the Liouville theory despite the fact that the classical
transport equation is linear (like the Schr\"{o}dinger equation or the
quantum Liouville equation): it arises simply because the delta
function or trajectory limit is allowed classically, but not quantum
mechanically.

\subsection{The Multiple Trajectory Gaussian Approximation}

The idea behind the MTGA is to improve the TGA by including multiple
classical trajectories. One does this by propagating more than one
Gaussian, using the TGA for each individual packet. There are two
reasons why this is a good idea. First, a sum of Gaussians can track
the classical distribution even at times when the distribution is
highly non-Gaussian (although this may involve a high computational
cost as very many Gaussians might be needed in the sum). The second
issue involves quantum coherence in phase space. The classical version
of this approximation simply sums over the individual Gaussians to
give the composite distribution function. This is not true in the
quantum case: a coherent sum of Gaussian wave functions, it turns out,
is not a sum of the associated Wigner functions, but rather a sum
augmented by ``off-diagonal'' quantum coherence
terms~\cite{HSC}. (These off-diagonal terms account for the Airy
fringes in the WKB approximation for the Wigner function discussed by
Berry~\cite{MB}.) Thus the MTGA contains nontrivial quantum
information, and is in this sense, unlike TDVP and TGA, not a
classical approximation.

\section{Summary}

The comparative evolution of c-number classical and quantum moments
yields an obvious method to compare classical and quantum mechanics as
dynamical theories. In doing so here, we have restricted attention
largely to the case of the Gaussian approximation. The time-dependent
Gaussian approximation in quantum mechanics has a direct, completely
equivalent, classical analog in terms of an ensemble of classical
particles also treated in the same approximation. Thus {\em every}
`quantum' result in this approximation has an exact classical
counterpart. Part of the reason for the misinterpretation of
`semiquantum' chaos as a quantum dynamical effect was simply due to a
too naive comparison with classical mechanics: a comparison against
trajectories, rather than against swarms of trajectories. Using the
more general approach presented here, it becomes clear that, in this
context, $\hbar$ plays no fundamental role. The key point is that a
dimensionful constant {\em also} appears in classical transport
theory: $\sigma_2$, as defined in Eqn.~(\ref{3b}), which plays the
role of an effective $\hbar$ there -- such a constant is missing from
the equations of motion of classical trajectories but arises once
classical distributions have to be evolved by the Liouville equation.

\section{Acknowledgements}

It is a pleasure to acknowledge helpful discussions (over a rather
extended period of time!) with Tanmoy Bhattacharya, Fred Cooper, Tim
Elston, Ronald Fox, Kurt Jacobs, Ronnie Mainieri, Peter Milonni,
Arjendu Pattanayak, Daniel Steck, Kosuke Shizume, Bala Sundaram, and
George Zaslavsky.

\end{document}